\documentclass[preprint2]{aastex}

\slugcomment{submitted to the Astrophysical Journal}

\shorttitle{RFI mitigation in radio Interferometric Data}
\shortauthors{Athreya}

\begin{document}


	\title{A new approach to mitigation of radio frequency interference in
interferometric data}


\author{Ramana Athreya}
\affil{National Centre for Radio Astrophysics,\\ P. O. Bag 3, Pune University
   Campus, Pune: 411 007. India; rathreya@ncra.tifr.res.in\\[4ex] Accepted for publication in the Astrophysical Journal}




\begin{abstract}
Radio frequency interference (RFI) is the principal factor limiting the
sensitivities of radio telescopes, particularly at frequencies below 1
GHz. I present a conceptually new approach to mitigation of RFI in
interferometric data. This has been used to develop a software tool 
(RfiX) to remove RFI from observations using the Giant Metrewave Radio
Telescope, India. However, the concept can be used to excise RFI in
any interferometer. Briefly, the fringe-stopped correlator output
of an interferometer baseline oscillates with the fringe-stop
period in the presence of RFI. RfiX works by identifying such a
pattern and subtracting it from the data. It is perhaps the only
purely software technique which can salvage the true visibility value
from RFI-corrupted data. It neither requires high-speed hardware nor 
real-time processing and works best on normal
correlator output integrated for 1--10s.  It complements other
mitigation schemes with its different approach and the regime it
addresses. Its ability to work with data integrated over many seconds
gives it an advantage while excising weak, persistent RFI unlike most
other techniques which use high-speed sampling to localise RFI in
time-frequency plane.  RfiX is also different in that it does not
require RFI-free data to identify corrupted sections.  Some results
from the application of RfiX is presented including an image at 240 
MHz with a Peak/noise ratio of 43000, the highest till date at 
wavelengths $>1$m.
\end{abstract}





\section{Introduction}

Radio Frequency Interference signals (RFI), which are generated by
human activities, are among the main factors limiting the performance
of radio telescopes.  RFI constricts the available frequency space,
effectively increases system noise and corrupts calibration solutions.
The effect is particularly strong at frequencies below 1 GHz.
Ironically, weak RFI cause more problems than strong RFI; the latter
are more easily identified and eliminated and the resulting loss of
data can be compensated by longer observations. In radio
interferometry \citep{thmosw} each complex visibility is measured by
averaging the output of a baseline over typically $1-30$ seconds.  RFI
fainter than the noise in each visibility will not be detected. During
the course of an observation the same baseline could contribute up to
tens of thousands of visibilites corrupted by the same RFI. It is
usually not possible to integrate out these correlated errors by
longer observations.

RFI removal schemes may be broadly classified into two categories: in
the first case the corrupted data is identified and flagged --- i.e.
it is completely lost. In the second case one measures the celestial
signal under the RFI, thus salvaging the data. The two categories have
not always been labeled differently in literature but increasingly
{\em excision} and {\em mitigation} have been used interchangeably in
literature to denote the latter category.

Most RFI removal techniques use high-speed sampling at nano- to
milli-seconds followed by a fourier transform to localise RFI in time
and/or frequency \citep{bhatnd, wks07, frid01, bfm04}.  Statistical
quantities of various degrees of sophistication \citep{frid08},
reference signals \citep{bb98} and the known properties of RFI signals
in a particular context \citep{bbk00, ebb01} have been used to
identify RFI.  Generally, these techniques work for the context they
were designed while none is universally applicable. Some of the
limitations are: (i) high-speed sampling techniques have higher
measurement noise and hence are less capable of identifying weak but
constantly present RFI, (ii) the requirement of clean reference areas to
identify RFI (iii) the necessity of specialised hardware. A different
class of techniques have used eigen decompositions of the correlation
matrices to isolate and excise RFI in interferometry \citep{lvb00,
kocz04, bv05, penetal08}. The reader is refered to the proceedings of
the RFI2004 workshop in Radio Science 2005, vol 40, and URSI 2008,
Chicago, for a compendium of RFI-related literature).

I describe here a new approach to mitigation of RFI from radio
interferometric data which can excise contributions from
quasi-constant sources of RFI. It was developed for analysing
low-frequency data from the Giant Metrewave Radio Telescope (GMRT),
Pune, India, and has been implemented as an offline software tool
({\em RfiX}). It complements extant techniques in approach and the
applicable RFI regime. In particular, a combination of RfiX and 
real-time high-speed sampled voltage clipping algorithms would go
a long way in minimising RFI in interferometers.
Since it works on the usual correlator output it can
also be used to process archival data.  However, the concept is
applicable to any interferometer which uses fringe-stopping.

\section{Fringe-Stop-Pattern for RFI mitigation} 
The RFI mitigation presented here is applicable to spatially and
temporally constant RFI sources. Moving RFI cannot be processed at all
and temporal variation will affect the efficacy of the algorithm.

In a standard radio interferometer the fringe pattern due to Earth's
rotation in the correlated output of a baseline is suppressed for the
phase centre in the sky by multiplying the two antenna data streams by
a (co)-sinusoid; i.e each baseline is fringe-stopped.
The fringe-stop frequency is given by 
\begin{eqnarray}
\label{fringefreq}
\nu_{_{F}}(t) = -\omega_{_{E}} \ U_{\lambda}(t) \ \cos{\delta(t)}
\end{eqnarray}
\noindent
where $U_{\lambda}(t)$ is the instantaneous standard spatial 
frequency component, $\delta$ is the declination of the phase centre,
and $\omega_{_{E}}$ is the Earth's angular speed \citep{thmosw}. In 
general, sources in
other locations will not be fringe-stopped and will contribute a
residual fringe to the baseline ... which, in fact, is used to
construct the sky intensity image. The residual fringe speed increases
with the radial distance of the source from the phase centre. Thus, a
source external to the target field will fringe faster than any
interior source along the same radial vector from the phase centre. In
particular, every
source of terrestrial RFI will fringe faster than any astronomical
source and, in fact, at exactly the fringe-stopping rate.

The last point may be understood by realising that a stationary RFI
source will contribute a constant amplitude and a phase to a baseline
which only depends on its location relative to the two antennas.
Fringe-stopping will cause the RFI signal to pick up the inverse of
the sinusoidal pattern. Thus, in a fringe-stopped interferometer the
observed modulation of a stationary RFI signal is independent of its
location on earth and only depends on the baseline length and
orientation relative to the phase centre (in the sky). Multiple RFI
sources will all be modulated by the same fringe-stop frequency.  All
such sources, even if they differ in amplitude and phase, will add
vectorially to contribute a single sinusoid to the baseline --- this
key point is worth emphasising. 

\begin{figure}
\epsscale{0.95}
\plotone{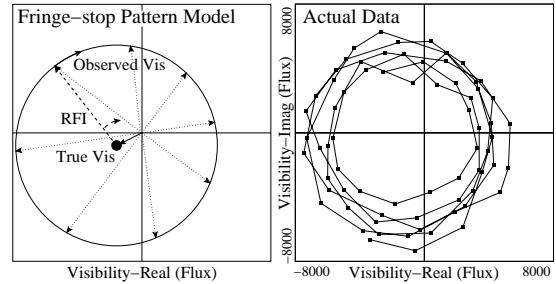}
\caption{\label{rfimod}
Effect of fringe-stopping and stationary RFI on the true visibility.
The rectilinear complex-plane plots (not to be confused with the UV
plane) show the temporal behaviour of the true and observed values of
the visibility from a single baseline.  Left: schematic illustration
--- fringe-stopping causes the RFI (dashed vector) to circulate around
the true visibility (solid vector) at the fringe-stop frequency for
the baseline. The resultant amplitude (dotted vectors) and phase
(their position angle w.r.t Real-axis) vary with time in an
analytically calculable manner (see Fig. \ref{amplphase}).  Right:
actual data --- 5 minute sequence of GMRT 153 MHz data from C00-C04
baseline exhibiting the visibility circulation. Adjacent points
along the curve are separated by 4.18 second.}
\end{figure}

\begin{figure}
\epsscale{0.95}
\plotone{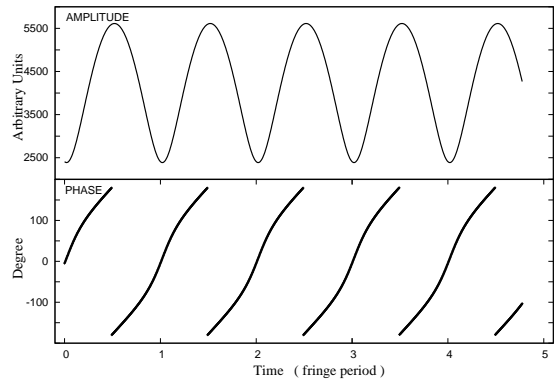}
\caption{\label{amplphase}
Observed visibility in the presence of RFI for the model shown in Fig.
\ref{rfimod}. The amplitude and phase, derived from Eqn \ref{vobs} are
plotted against time. The observed data may be fit by well-defined
analytical expressions to estimate the contribution of RFI to the 
observed visibility.} 
\end{figure}

The combined effect of RFI and fringe-stopping is shown
schematically in Fig. \ref{rfimod} along with actual data; the observed
visibility will then be given by:
\begin{eqnarray}
\label{vobs}
V_{_{OBS}} = V_{_{TRUE}} + \mathcal{A}\ e^{i[{2\pi\nu_{_{F}}(t)}t -
\Phi]} + Noise
\end{eqnarray}
\noindent
where, all terms are complex, $\mathcal{A}$ and $\Phi$ are the
amplitude and phase of the RFI in the baseline and $\nu_{_F}$ is the
fringe-stop frequency (Eqn. \ref{fringefreq}). Provided the RFI remains
constant over a good fraction of the fringe period, the observed data
may be fitted for fit-average values of $\mathcal{A}$, $\Phi$, and
$V_{_{TRUE}}$. The observed visibility amplitude and phase obtained 
in a closed/analytical form from Eqn. \ref{vobs} are plotted in Fig.
\ref{amplphase}. The salvaged/RFI-free visibility will then be
\begin{eqnarray}
\label{vrfix}
V^j_{_{RFIX}}  =  V^j_{_{OBS}} - 
        \mathcal{A}e^{i[{2\pi\nu_{_{F}}(t)}t^j-\Phi]} 
                \equiv V_{_{TRUE}} + Noise^j
\end{eqnarray}

\noindent
where, $j$ is an index identifying a particular data point in the
subset over which the RFI was presumed to be constant and a model was
fit.  Note that $V_{_{TRUE}}$, $\mathcal{A}$ and $\Phi$ do not carry
the index $j$ and are average fit values for the entire subset. The
average $V_{_{TRUE}}$ obtained from the fit is not carried to the next
stage.  Instead, the model RFI contribution ($\mathcal{A}$ and $\Phi$)
for the fitted subset is subtracted from the j-th observed visibility
(within the subset) to obtain the j-th modified visibility; i.e. the
modified values continue to carry the original unsmoothed noise
component and hence adjacent values remain independent.

This technique does not affect the all-important closure relationships
\citep{thmosw}. The mean value of $V_{_{OBS}}$ (i.e. the centre of the
circle) is a good estimator of $V_{_{TRUE}}$ and can be used to
quantify deviations from closure, if any.  Alternatively, one can
consider the case of an inaccurate estimate of the RFI amplitude and
phase. Since the modified visibilities are obtained by subtracting the
RFI circle, an inaccurate RFI model is equivalent to the original RFI
corruption being replaced by a new RFI corruption, whose amplitude
should be close to zero under ideal conditions but in any case smaller
than the original RFI amplitude. In practice, a simple
before-and-after rms calculation and $\langle V_{_{RFIX}}\rangle -
\langle V_{_{OBS}}\rangle$ can be used to quantify closure deviations,
if any, and provide the choice of either using the model fit or
retaining the original data --- at worst the processed data is no
worse than the raw data. 

The relative strengths of the RFI and true visibility is not relevant
either since the algorithm is being applied on the rectilinear complex
plane.

\section{Implementation at GMRT --- RfiX package}

The RFI environment at GMRT can be quite intimidating, especially at
frequencies below 400 MHz. While the strongest source in a field is
typically only a few Jy, the RFI can occasionally reach several
hundred Jy in some channels in the 150 and 240 MHz bands.

However broadband RFI of up to several tens of Jy
causes even more damage: currently available tools, including the
Astronomical Image Processing System (AIPS) and FLAGIT (a locally
developed tool) would not even recognise the RFI contamination ---
there is no "clean" reference area for identification by contrast.

\begin{figure}
\epsscale{0.95}
\plotone{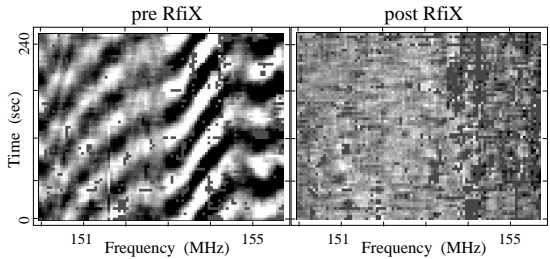}
\caption{\label{timechan}Pre- and post-RfiX views of the visibility
Time-Channel plane. Each pixel is 4.18 second in time and 62.5 KHz in
frequency. Every single channel is afflicted by RFI and each different
gradiant seen on the left panel represents a different RFI source.
Application of RfiX results in a more-or-less featureless, i.e.
RFI-free, plane (right). RfiX does not require clean reference areas to
eliminate RFI.}
\end{figure}

This may be seen in Fig. \ref{timechan} which plots the visibilities
in grey-scale on a 2-dimensional time-frequency plot. In the absence
of RFI all the visibilities in a small subsection of the
time-frequency plane should more-or-less represent the same fourier
component for a continuum source and hence the intensity pattern
should be devoid of features. A very narrow RFI line would show up as
a ripple along the time axis in a single frequency channel. A coherent
broad-band RFI source would introduce a ripple across many channels
and the direction of the pattern would depend on the location of the
RFI with respect to the baseline. The ripples seen in the left-hand
panel show that the visibility data shown has RFI in every single
time-frequency pixel. The different slopes seen in the
pattern points to the presence of multiple sources of RFI which have
emission broad enough to cover many channels (of up to several MHz)
but not broad enough to cover the whole band, which complicates data
analysis as will be described shortly.

Broadband RFI causes calibration errors which severely limit the
dynamic range (Peak-to-rms ratio) to less than a few thousand at these
frequencies, even for observations of several full synthesis sessions
(each of 8-10 hours). Even though the
GMRT is currently the most sensitive low frequency instrument the
sensitivity achieved at 150 and 240 MHz is more than a factor of 10
poorer than expected for long integrations.

There are two ways of removing RFI from the data. A fourier transform
of the time-frequency visibility data would efficiently localise the
RFI signal in frequency-lag space (time-lag $\equiv$ path difference
from the RFI source to the 2 antennas).  A fourier transform requires
many dozens or even hundreds of data points within the period of
constancy of the RFI source for efficient inversion. This requires
higher (than usual) speed sampling and storage for consequent 
larger data volume.
Furthermore, abrupt changes along either the time- or the
frequency-axis, like the ones seen in Fig. \ref{timechan}, will make
the frequency-lag plane noisier, reducing the ability to detect RFI.

The alternative chosen for RfiX was motivated by the realisation that
while RFI ripples along the frequency axis could have different
scale-lengths they would all have exactly the same period along the
time axis and this period was identical to the analytically defined
fringe-stop period used by the correlator. The RFI in each channel was
estimated independently by fitting a 1-dim sinusoid in time (instead
of a 2-dim fourier transform), which has several advantages: (i) it
eliminates all modulation scales other than the fringe-stop period
from the analysis; fitting a single sinusoid of known period is a
simple linear process with well-defined error estimates, (ii) it
affords greater flexibility and efficiency in identifying the time
subsections of constancy of RFI in each channel, thus accommodating
the abrupt changes in the time-frequency plane seen in Fig.
\ref{timechan}, and (iii) since the model only has 4 free parameters
and the sinusoid period is fixed one can safely restrict the fitting
process to small data sections and hence greater chance of constancy
of RFI and more accurate RFI excision.

While the standard mode of the GMRT correlator outputs visibilities
every 16.7s one can also trivially select faster data output modes
which provide visibilities every 2 or 4, and exceptionally even 0.128,
seconds. Recording the data every $2-4$ seconds is sufficient for this 
algorithm; one should compare this
to the nano- to millisecond sampling rate usually required by most
other techniques of RFI removal.
Using data integrated for several seconds has the advantage of (i) not
requiring additional hardware (ii) easily
manageable data output rate and hence (iii) the option of offline
processing. Additionally, this longish integration time reduced the 
measurement noise in each visibility and allowed for detection and
excision of fainter RFI.

\begin{figure}
\epsscale{0.95}
\plotone{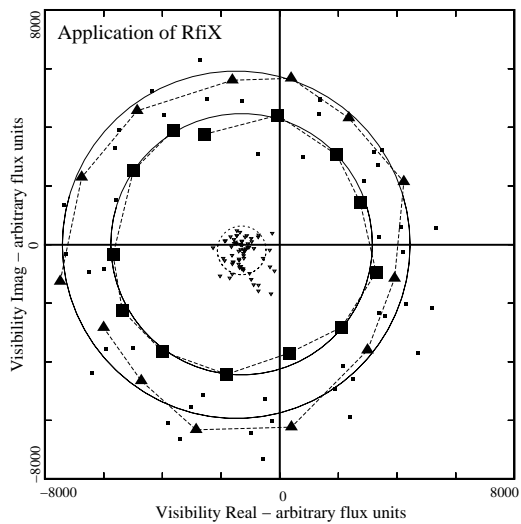}
\caption{\label{rfixapln}
RfiX algorithm, applied to the observed data plotted in Fig. \ref{rfimod}.
The outer annulus of points (large squares, triangles and dots) is the raw
data from a 5-minute GMRT scan at 153 MHz.
The two subsets of 13 data points (large squares and triangles) connected
by dashed lines are two temporal sequences spanning one fringe period 
each and the solid circle associated with each is the model fringe due to
RFI. The other subsets and their fits have been omitted for clarity. The 
inner cluster of points are the modified (true) visibilities obtained by 
subtracting the model fringe from the raw visibilities. The small dashed
circle around the inner cluster denotes the 3$\sigma$ of the scatter.} 
\end{figure}

Figure \ref{rfixapln} illustrates the application of
the algorithm to actual data.  The plot shows a 5 minute data sequence
from a single channel (same as in Fig. \ref{rfimod}).  The outer
annulus of points is the raw data. The compact cluster at the centre
is the processed/RFI excised data. The variability of RFI within the 5
minute scan was surmounted by separate fits to subsets of the data of
length 54s (= 1 fringe period). Nevertheless, this variability
resulted in imperfect RFI mitigation, which is reflected in a larger
scatter in the modified visibilites.  However, RfiX succeeded in
reducing the scatter by a factor of 5, from the radius of the outer
annulus (RFI amplitude) to the post-processing scatter represented by
the small inner circle. This much smaller scatter facilitates further
cleaning-up through sigma-clipping. 

The fully automated implementation: (i) checks if the baseline 
fringe-stop period is within the appropriate range (see under 
Discussion) and if not, either skips it or flags it as specified (ii)
compares the scan rms to a fiducial value to determine if the
particular channel has significant quantum of RFI and therefore worth
processing, (iii) identifies subsections with constant RFI amplitude 
and (iv) robustly fits and subtracts the RFI fringe

\begin{figure}
\epsscale{0.95}
\plotone{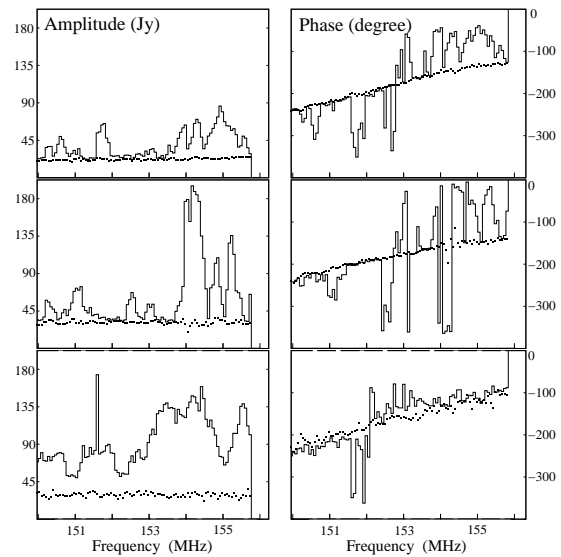}
\caption{\label{scanap}Efficacy of RfiX in cleaning the bandpass.
The panels show the average bandpass for three 5-minute scans spanning 
a 4-hr period of the strong calibrator 3C446 at 153 MHz using the GMRT.
The solid histograms are the pre-RfiX (raw) data and the
dots are the post-RfiX visibilities. The source flux density is 21.5
Jy. The strongest RFI (middle scan) is almost 200 Jy. The bottom scan
has at least 25 Jy of RFI in every channel and four times as much in
some. In spite of the very strong RFI the post-RfiX bandpass is more or
less flat in amplitude and linear in phase and the amplitude level and
phase gradient is similar in all 3 scans.}
\end{figure}

Figure \ref{scanap} shows the results of RfiX on the averaged bandpass
of amplitude and phase of three different 5-minute scans (separated by
4 hours).  The data was obtained with the GMRT at 153 MHz on the
calibrator source 3C446 (21.5 Jy). The upper panel shows RFI nudging
200 Jy in some channels. The lowermost panel is a remarkable example
of the algorithm's efficacy. The entire band is corrupted by at least
25 Jy of RFI, with even larger contamination in some channels.
Traditional methods of offline RFI clipping (in AIPS, for instance)
would completely fail in this case. The large amplitude bias would
have cascaded across the data during self-calibration, vitiating the
gain solutions of all antennas including those without RFI.  While
this is an extreme example it happens often enough to seriously limit
the sensitivity at 130-250 MHz. Even though each frequency channel in
each scan was processed independently and the RFI varied considerably
the amplitude is more-or-less flat across the band and from scan to
scan, as also the phase gradient across the band. This is good
evidence that the algorithm is not corrupting the closure quantities. 

Figure \ref{timechan} shows the visibility time-channel plot of a scan
and the efficacy of RfiX even when every single channel and time is
affected by strong RFI.

\begin{figure*}
\epsscale{1.95}
\plotone{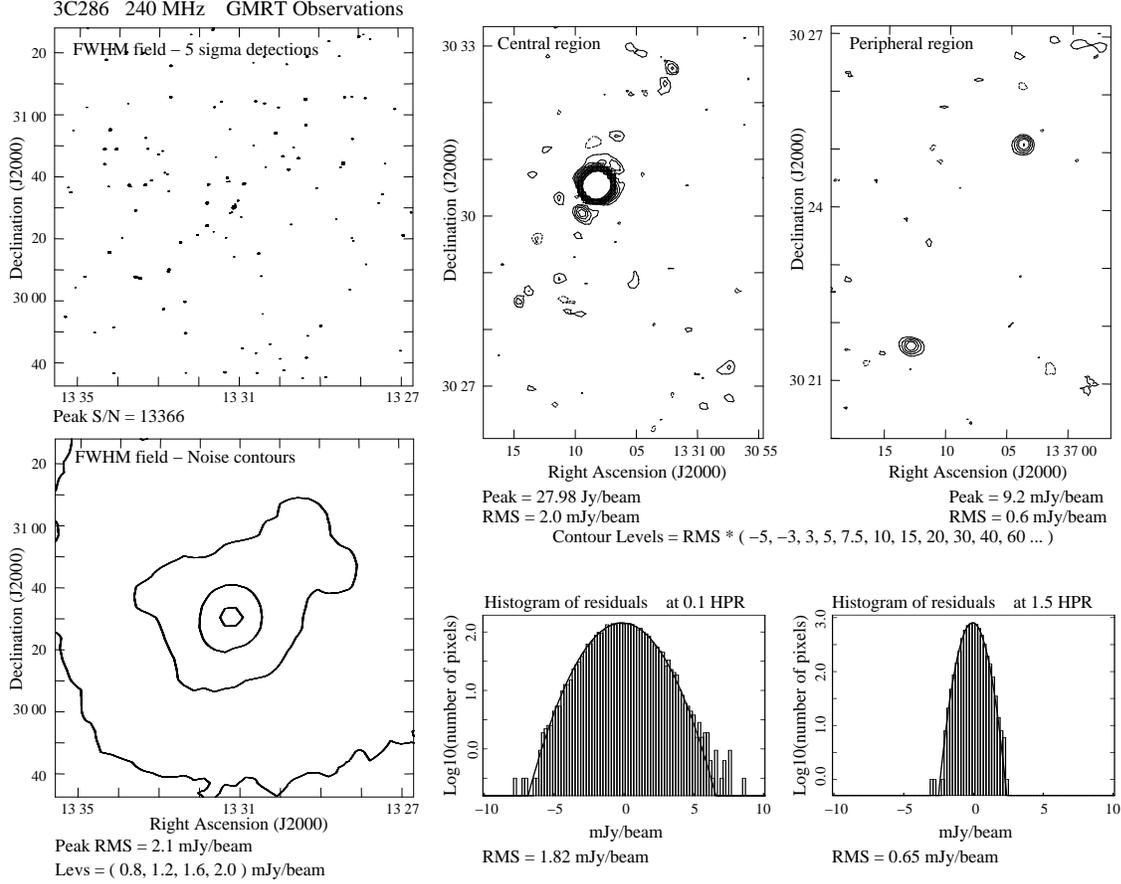}
\caption{\label{fwhmimage}GMRT image of 3C286 at 240 MHz using RfiX
processed data.
The left panels show the celestial sources (top) and the rms noise 
(bottom), respectively. They were made using just 2 hours of 
data with the GMRT. The field covers the half-power extent of the
primary beam. The peak flux density is 28 Jy (of 3C286) and the 
noise is 2 mJy/beam near 3C286 and 0.65 mJy/beam at
the periphery ($Peak/noise \equiv$ centre: 14000, periphery: 43000).
This is perhaps the highest dynamic range image made at wavelength $>$
1m (frequency $<$ 300 MHz).  The plot illustrates, rather starkly, the
problem of finding isolated point sources for calibration at low
frequencies!
The top-row middle panel shows a more detailed plot of the immediate
area around 3C286. As may be expected artifacts are most
numerous here, but even so the image is quite clean with only few 
artifacts in excess of $5\sigma$, with the highest at $8\sigma$. In the
peripheral regions (top-right) there are no artifacts in excess of 
$4\sigma$. Histograms of the noise pixels (bottom row) near the contour
locations show well-behaved residuals.}
\end{figure*}

Figure \ref{fwhmimage} shows an image of the 3C286 field from 2 hours
of GMRT observations at 240 MHz. 3C286 is 28 Jy at that frequency.
The rms noise at the centre of the field (close to 3C286) is
2 mJy/beam ($Peak/noise \sim 14000$). Furthermore, despite the very
strong source the central region is mostly clean with only one
artifact reaching 8$\sigma$. The noise at the periphery is only 0.65
mJy/beam ($Peak/noise \sim 14000$). Histograms of the noise pixels
show that the image is "clean" with no residuals in excess of
4$\sigma$. This is perhaps the highest dynamic range image achieved
till date at frequencies below 300 MHz (wavelength $>$1m).

\section{Discussion}

The most important feature of RfiX is its ability to salvage
RFI-corrupted data --- it not only identifies RFI but also excises it
to obtain uncorrupted data. This is perhaps the only RFI mitigation
tool which is entirely software based.
It can be applied offline on the normal correlator data
typically output every $0.1-30$ seconds. This much longer integration
compared to high-speed sampling techniques makes RfiX sensitive to
fainter RFI. It complements high-speed techniques which are best for
clipping short but intense bursts of RFI but cannot target weak but
constantly present RFI. Also, RfiX does not require a clean reference
area to identify RFI. 

The technique is applicable to any interferometer which uses
fringe-stopping. It can also be applied at any frequency provided 
only that the correlator output matches the Nyquist rate for the
fringe-stop frequency of the particular baseline. For instance, the
GMRT correlator outputs data every 2-16
seconds in the standard mode. A 2s sampling rate limits the 
application of RfiX to baselines with $U_{\lambda} < 3-12 k\lambda$ 
(declination range $0^o-75^o$). At the GMRT the RFI is most severe at 
$\nu_{OBS} < 400$ MHz and at these frequencies the above limit
corresponds to baselines of several kilometres. We do not usually
find correlated RFI on such long baselines and so the sampling rate
does not pose a constraint in practical terms. In any case, the 
present correlator at GMRT can be configured to provide data every 
0.128 second and so, if necessary, the longer baselines and/or higher 
frequency observations can also be processed for RFI using RfiX. In
general, a visibility sampling interval of 0.1s, which can be 
very easily achieved using present-day hardware, will be sufficient
for excising RFI (using RfiX) from baselines up to a few km in 
observations up to a few GHz. Of course, shorter visibility
integrations in the correlator can relax both these limits.

RfiX will not work effectively with interferometric observations near
the Celestial Poles where the fringe rate is very slow. Even in
regions away from the Poles baselines with small $U$-component (see
Eqn. \ref{fringefreq}) have very long fringe periods and so cannot be
satisfactorily processed. RfiX works best for RFI which manifests and
remains constant over the fringe period of a baseline. Fitting over a
full fringe period is very safe for conserving closure. Fitting
intervals shorter than a third of the fringe period could lead to
incorrect estimation of RFI amplitude and hence closure-violating
baseline errors. At the GMRT the RFI amplitude can be quasi-constant
for up to 150-200 seconds which limits the application of
RfiX to baselines with fringe periods $\leq500$ seconds, corresponding
to baseline component $U_{\lambda}$ $\geq25-100\lambda$ (declination 
range $0^o-75^o$). Of course, we do see instances of RFI varying on
much shorter timescales and RfiX will fail in those cases. Depending 
on the goal one can either flag all shorter baselines or retain them 
without RfiX processing.  Therefore, this algorithm will not improve the
quality of image structures larger than $0.^{\rm{o}}5-1.^{\rm{o}}5$.
This is not a serious limitation since the GMRT primary beam FWHM is
not much larger. Note that the constraint comes from the timescale of
RFI variability and not RfiX. \cite{penetal08} also have a similar
limitation for excising RFI in baselines with long fringe periods.

Briefly, the limiting fringe stopping frequency and the corresponding
baseline for which RfiX is effective are related through the rather
simple Equation \ref{fringefreq}. It must be noted that only the
baseline component $U_{\lambda}$ is involved in the equation
and not $V_{\lambda}$. The lower baseline limit is
defined by the timescale of RFI variability while the upper limit is
set by the visibility integration period of the correlator.

It is often claimed that interferometric fringe-stopping in itself
washes out RFI, but this is not entirely appropriate for low frequency
arrays. If the system temperature (T$_{\rm{sys}}$) dominates over the
RFI amplitude (A$_{\rm{RFI}}$) the noise in a single visibility of
integration period $\tau$ will be T$_{sys}/\sqrt{\tau}$. When RFI
dominates the effective noise will be A$_{\rm{RFI}}/(\nu_{_F}\tau)$,
where $\nu_{_F}$ is the fringe-stop frequency for the particular
baseline.  Thus, RFI suppression due to fringe-stopping is most
effective for short baselines for which $\nu_{_F}\tau \gg 1$. In the
context of GMRT observations at 153 MHz (longest baseline 27 km) the
shortest fringe period is of the order 1 second (see Eqn.
\ref{fringefreq}) compared to the recommended integration time of 2-4
seconds. Therefore RFI suppression due to fringe stopping would only
have a marginal impact.  Even at 1 GHz RFI would be suppressed only in
baselines formed by the outer antennas; but correlated RFI is anyway
mostly limited to baselines shorter than a few km. RFI suppression due
to fringe-stopping would be most effective in arrays operating at $\nu
\gg 1$ GHz as even the shorter (RFI-prone) baselines will have 
sufficiently short fringe period to achieve RFI suppression.

A detailed description of the implementation of RfiX for GMRT data is 
presented in \citep{A09} and may be of use to astronomers intending to
implement the tool at other facilities.

RfiX is somewhat related to the RFI-to-North-Pole scheme tried by
\cite{rfinp}.  Briefly, a stationary RFI and a source at the
North-Pole are both stationary with respect to the
interferometer. Thus, one can attempt to map RFI to pseudo sources at
the North Pole and subtract them from visbilities. This will work when
the RFI contributes the same antenna temperature to all the antennas,
just like a celestial source. However, in general the intensity
of an RFI source varies from antenna to antenna due to $1/R^2$
fall-off and shielding by manmade or geographical structures. RfiX
surmounts this problem by processing each baseline separately allowing
for variable intensity in each.

In summary, RfiX is a powerful tool based on a simple concept for
mitigating RFI in interferometric data. Indeed, its strength lies in
its simplicity which provides a clear understanding of its domain of
applicability. It not only identifies RFI but also salvages
RFI-affected visibilities. It is a purely software approach which
requires neither additional hardware, nor real-time analysis nor
complicated mathematical operations and can even be applied to
archival data. It complements other extant excision techniques in
approach and domain of applicability. It will work for any
interferometer which uses fringe-stopping. 

\acknowledgments
I thank Pramesh Rao and Dipanjan Mitra for useful discussions 
during the protracted incubation of this algorithm, and an anonymous 
referee for useful comments. K.S.Jeeva
contributed data display and FITS IO tools. Rahul Basu's work helped to
flag residual RFI from 3C286 data. I thank the GMRT staff for 
helping with observations presented in this work.

{\it Facilities:} \facility{GMRT}.


\clearpage

\begin{thebibliography}{}

\bibitem[Athreya 2009]{A09} Athreya, R. M. 2009, National Centre for 
   Radio Astrophysics Tech. Report, in prep.
\bibitem[Baan et al. 2004]{bfm04} Baan, W. A., Fridman, P.
   A., Millenaar, R. P.  2004, \aj, 128, 933
\bibitem[Bhat et al 2005]{bhatnd} Bhat, N.D.R., Cordes, J.M., Chatterjee, S.,
   Lazio, T.J.W.  2005, Radio Science, 40(5) (arXiv:astro-ph/0502149v1)
\bibitem[Boonstra \& van der Tol 2005]{bv05} Boonstra, A. J., van der Tol, S. 
   2005, Proc of the RFI2004 workshop, Radio Science, 40
\bibitem[Bradley \& Barnbaum 1998]{bb98} Bradley, R., Barnbaum, C.  1998, 
   \aj, 116, 2598
\bibitem[Briggs et al. 2000]{bbk00} Briggs, F. H., Bell, J. F., Kesteven,
   M. J.  2000, AJ, 120, 3351
\bibitem[Ellingson et al. 2001]{ebb01} Ellingson, S. W., Bunton, 
   J. D., Bell, J. F.  2001, \apjs, 135, 87–93.
\bibitem[Fridman 2001]{frid01} Fridman, P.A.: 2001, \aa, 368, 369
\bibitem[Fridman 2008]{frid08} Fridman, P.A.: 2008, \aj, 135, 1810
\bibitem[Kocz 2004]{kocz04} Kocz, J. 2004, Radio frequency interference 
   characterisation and excision in radio astronomy, Honours thesis, Dept of
   Engineering, The Australian National University.
\bibitem[Lesham et al. 2000]{lvb00} Leshem, A., van der 
   Veen, A. J., Boonstra A. J.  2000, \apjs, 131, 355
\bibitem[Pen et al. 2008]{penetal08} Pen, U.-L., Chang, T.-C., Peterson, 
   J. B., Roy, J., Gupta, Y., Bandura, K.  2008, arXiv:astro-ph/0804.2501v1
\bibitem[Perley et al. 2005]{rfinp} Perley, R. A., Cornwell, T. J., Bhatnagar, 
   S., Golap, K.  2005, Proc of the RFI2004 workshop, Radio Science, 40
\bibitem[Thomson et al. 1986]{thmosw} Thomson, A. R.,
   Moran, J. M., \& Swenson G. W.    1986, 
   Interferometry and Synthesis in Radio Astronomy (Wiley Interscience, 
   New York). 
\bibitem[Winkel et al. 2007]{wks07} Winkel, B.,
   Kerp, J., \& Stanko S.   2007, Astron. Nachr., 328, 68
\end{thebibliography}
\end{document}